\documentclass[12pt]{article}
\usepackage{epsfig}
\usepackage{hyperref}

\usepackage{parskip}

\usepackage{color} 
\definecolor{rougef}{rgb}{0.7,0,0}
\definecolor{vertf}{rgb}{0,0.6,0}
\definecolor{bleuf}{rgb}{0,0,0.9}

\newcommand{\beq}{\begin{equation}}
\newcommand{\eeq}{\end{equation}}
\newcommand{\beqa}{\begin{eqnarray}}
\newcommand{\eeqa}{\end{eqnarray}}
\newcommand{\beqar}{\begin{eqnarray*}}
\newcommand{\eeqar}{\end{eqnarray*}}
\newcommand\cT{T}

\newcommand{\al}{\alpha}
\newcommand{\be}{\beta}

\def\spa          {\ \ \ }

\def\ha           {\mbox{$\frac{1}{2}$}}

\def\spa          {\ \ \ }

\def\Tr           {\mbox{\rm Tr}\,}
\def\STr          {\mbox{\rm STr}\,}

\def\cd           {{\cdot}}
\def\ran          {\rangle}
\def\lan          {\langle}
\def\fsH	{H\!\!\!\!/\,}

\newcommand{\eps}{\epsilon}
\newcommand{\ga}{\gamma}

\newcommand{\inn}{\!\cdot\!}
\newcommand{\h}{\eta}

\newcommand{\lam}{\lambda}

\newcommand{\z}{\zeta}

\newcommand{\ie}{{\it i.e.,}\ }
\newcommand{\labell}[1]{\label{#1}} 
\newcommand{\reef}[1]{(\ref{#1})}
\newcommand\prt{\partial}

\newcommand\cL{{\cal L}}

\newcommand\bD{\bar{D}}

\usepackage[utf8]{inputenc}

\begin{document}

\begin{center}
\thispagestyle{empty}

{\huge All-Order Quartic Couplings in  Highly \\[10pt] Symmetric D-brane-Anti-D-brane Systems}

\vskip 1cm

{ Ehsan Hatefi}${}^{a,}$\footnote{\href{mailto:ehsan.hatefi@sns.it, ehsanhatefi@gmail.com}{\texttt{ehsan.hatefi@sns.it, ehsanhatefi@gmail.com}}} and  { Per Sundell}${}^{b,}$\footnote{
\href{mailto:per.sundell@unab.cl}{\texttt{per.sundell@unab.cl, per.anders.sundell@gmail.com}}}\\[3pt]

\vskip .4cm

$^{a}$ \textit{Scuola Normale Superiore and I.N.F.N}\\
\textit{Piazza dei Cavalieri 7, 56126, Pisa, Italy
}
  
$^{b}$ \textit{Departamento de Ciencias F\'isicas, Universidad Andres Bello}\\ 
\textit{Sazi\'e 2212, Santiago, Chile}

\end{center}
\vskip 1cm
\paragraph{Abstract:}  We compute six-point string amplitudes for the scattering of one closed string Ramond-Ramond state, two tachyons and two gauge fields in the worldvolume of D-brane-anti-D-brane systems in the Type II superstring theories.
From the resulting S-matrix elements, we read off the precise form of the couplings, including their exact numerical coefficients, of two tachyons and two gauge fields in the corresponding highly symmetric effective field eheory (EFT) Lagrangian in the worldvolume of D-brane-Anti-D-brane to all orders in $\alpha'$, which modify and complete previous proposals.
We verify that the EFT reproduces the infinite collection of stringy gauge field singularities in dual channels.
Inspired by interesting similarities between the all-order highly symmetric EFTs and holographic duals of Vasiliev's higher spin gravities \`a l\`a Nilsson and Vasiliev, we make a proposal for tensionless limits of D-brane-anti-D-brane systems.
 
\vskip 7cm

\newpage

\clearpage
\setcounter{page}{1}

\tableofcontents

\section{Introduction}

In this paper, we shall study the effects of stringy corrections to gauge theory and gravity by extracting effective field theory (EFT) actions involving such corrections to all orders alias all-order EFTs.
As these theories reproduce the string amplitudes with their characteristic channel dualities, they are also sometimes referred to as highly symmetric EFTs.
More precisely, we shall focus our attention to all-order EFTs on worldvolumes of  D-brane-anti-D-brane (D$\overline{\rm D}$) systems in which all stringy modes have been integrated out save massless gauge modes and tachyons.
These models have a number of interesting applications ranging from the study of tachyon condensation in string theory via open-closed string duality to holography in non-BPS backgrounds and in high-energy limits.

String theory provides an ultraviolet completion of gravity by systematically adding higher-derivative corrections to its low-energy approximation given by Einstein's theory. 
Its power relies on the the fact that it intertwines two quantum theories in a nontrivial fashion such that the stringy corrections to gravity can be organized into a double perturbative expansion.
Thus, viewing gravity as a quantum field theory with an S-matrix, its stringy quantum corrections appear weighted by a string coupling constant $g_s$, which one may thus identify as $\hbar$, while at each order in $g_s$, the S-matrix elements are computed using a first-quantized string described by sigma model on a two-dimensional surface, alias worldsheet, of fixed genus, with its own coupling constant, often denoted by $\alpha'$. 
Physically speaking, $\alpha'$ sets the unit of length for the spacetime background metric, which yields an identification of the Planck length as a combination of positive powers of $g_s$ and $\alpha'$, which is arguably one of the most exciting results of string theory.

Another remarkable feature of the double perturbative expansion, is that it admits a natural extension so as to include non-perturbative effects by means of geometric objects permeating the spacetime background, known as Dirichlet-branes, or D-branes.
These may be thought of in two dual fashions as either defects in spacetime, serving as sources for stringy analogs of electric and
magnetic fluxes, or defects on the worldsheet, serving as sources for modified boundary conditions arising by means of two-dimensional analogs of electric-magnetic duality rotations.

In this paper, we shall attempt to attain a deeper understanding of the $\alpha'$-corrections to gravity and obtain a more complete control of their all-order behavior by focusing on  D$\overline{\rm D}$ systems.
The basic motivation behind studying these systems is that the aforementioned stringy dualities have so far been spelled out mainly in perturbative expansions around supersymmetric string vacua, while supersymmetry is broken throughout the bulk of the moduli space of string theory. 
To explore non-supersymmetric string backgrounds, D$\overline{\rm D}$ systems are important tools as their underlying first-quantized descriptions remain tractable and amenable to quantitative studies even in the absence of supersymmetry.  The detailed link between closed-string and open-string constructions originates from  \cite{Sagnotti}.

More broadly, in view of work on instabilities, and in particular on the importance of non-BPS branes in the Swampland \cite{Vafa:2005ui} and the related gravity conjecture \footnote {This conjecture confirms the presence of the light elementary electric and magnetic objects so that their mass/charge ratio is smaller than the relevant ratio for macroscopic extremal black holes and hence allows the decay of extremal black holes. } \cite{ArkaniHamed:2006dz}, one would like to gather additional insights into supersymmetry breaking on time--dependent backgrounds \cite{Gutperle:2002ai,Lambert:2003zr,Sen:2004nf}. 
For instance, if we use the Sakai--Sugimoto model of D$\overline{\rm D}$ systems \cite{Sakai:2004cn} to analyse spontaneous supersymmetry breaking in various holographic models \cite{Casero:2007ae}, the model will be granted as a probe if and only if the number of flavours is much less than the number of colours, $N_f<< N_c$. 
Hence, we shall keep all tachyonic string states in both worldsheet and EFT descriptions of the D$\overline{\rm D}$ system.

In \cite{Sen:1999md,Bergshoeff:2000dq}, various authors have found tachyonic effective actions that also describe the decays of non-BPS branes \cite{Sen:2002in}. 
Subsequently, in \cite{Sen:2004nf}, it was argued how to embed non-BPS branes and massless states into the EFT, whereby tachyon condensation for brane-anti-brane systems can be discussed in great detail \cite{Sen:1998sm}.

In the case of a single D-brane and a single anti-D-brane separated by a distance below the string length, two real tachyon modes emerge in the EFT.
In \cite{Michel:2014lva}, the dynamics of this system was analyzed using EFT methods, taking into account string loop divergences. 
As applications of tachyonic EFTs, we would like to highlight brane production \cite{Bergman:1998xv} and inflation in string theory, which later on became important ingredients the context of KKLT \cite{Dvali:1998pa}.
Various thermodynamic aspects of this system at finite temperature have been studied, whereby the system gets stabilised and can be related to black holes, leading to applications in AdS/CFT correspondence in string and M-theory \cite{Hatefi:2012bp}. 
Properties of D$\overline{\rm D}$ systems facilitate stability analyses in the context of KKLT and the Large Volume Scenario of string compactifications \cite{Polchinski:2015bea}. 
Including D-brane sources \cite{Polchinski:1995mt}, one may furthermore consider D$\overline{\rm D}$ bound states within the context of K-theory \cite{Witten:1995im}. 

In the aforementioned context, where the dualities of the supersymmetric M-theory web do not hold anymore, worldsheet Conformal Field Theory (CFT) methods are nonetheless still available, and remarkably they suffice for constructing string corrections and new couplings \cite{Myers:1999ps,Bjerrum-Bohr:2014qwa}.
Proceeding with this program, all standard EFT methods of effective actions for both non-BPS and BPS branes have been clarified with full details in \cite{Hatefi:2010ik}.
More precisely, the string corrections to DBI, Chern-Simons, Wess-Zumino effective actions are derived from scattering amplitudes, as, to the best of our knowledge, at least in the context of non-BPS branes, this approach does not introduce any ambiguities into $\alpha'$-corrections, and one explores all-order coefficients of string corrections in the Type IIB and IIA superstrings accordingly.

Finally, in the Conclusions we point to interesting similarities between the all-order EFT Lagrangian for D$\overline{\rm D}$ systems and the systems of matter fields coupled to topological conformal higher spin gravities that have been proposed \cite{Vasiliev:2012vf,Nilsson:2013tva,Nilsson:2015pua} as duals of bulk higher spin gravity a la Vasiliev \cite{Vasiliev:1992av} in an attempt to refine the original holography proposal \cite{HaggiMani:2000ru,Sundborg:2000wp,Sezgin:2002rt,Klebanov:2002ja}.
Motivated by this, within the context of the Frobenius--Chern--Simons (FCS) extension  \cite{Boulanger:2015kfa} of Vasiliev's higher spin gravity, we propose that the counterparts of D$\overline{\rm D}$ systems are contained as branches of fractional spin gravities \cite{Boulanger:2013naa}.

The paper is structured as follows:
We first concentrate on the D$\overline{\rm D}$ system and explain briefly the details on four- \cite{Kennedy:1999nn} and five-point functions.
We then explore more hidden symmetries in amplitudes of non-BPS branes.  
Having taken the selection rules \cite{Hatefi:2013yxa}, EFT and other symmetries into account, we explore the expansion of the amplitudes and start to exhibit all singularities of the type IIA and IIB string theories. 
This leads to the construction of the all-order $\alpha'$ higher-derivative corrections to two-tachyon and two-gauge field couplings in the D$\overline{\rm D}$ system as well.

This result shows that not only the structures of the higher derivative corrections but also their coefficients are different from the ones of non-BPS branes, as we shall clarify in detail in section five.

\section{Highly Symmetric D-Brane-Anti-D-Brane EFT}

The D$_p\overline{\rm D}_p$-brane system is constructed by $(-1)^{F_L}$-projecting the two branes which yields an unstable configuration containing tachyonic states. 
The resulting EFT of tachyons and massless gauge fields is formulated in terms of a superconnection of a noncommutative geometry ~\cite{quil,berl,Roepstorff:1998vh}, viz.
\begin{displaymath}
i{\cal A} := \left(
\begin{array}{cc}
  iA^{(1)} & \beta T^* \\ \beta T &   iA^{(2)} 
\end{array}
\right) \ ,
\end{displaymath}
where $A^{(1)}$ and $A^{(2)}$ are gauge fields associated to the D$_p$- and $\overline{\rm D}_p$-brane, respectively, and $T=\frac1{\sqrt{2}}(T_1+iT_2)$ is a complex tachyon with normalization $\beta$.
The resulting supercurvature
\beqa {\cal F}&:=&d{\cal A}-i{\cal A}\wedge\cal A=\left(
\begin{array}{cc}
iF^{(1)} -\beta^2 |T|^2 & \beta (DT)^* \\
\beta DT & iF^{(2)} -\beta^2|T|^2 
\end{array}
\right) \ , \eeqa
where $F^{(i)}:=dA^{(i)}-i A^{(i)}\wedge A^{(i)}=\frac{1}{2}F^{(i)}_{ab}dx^{a}\wedge dx^{b}$ and $DT:=dT-i(A^{(1)}T-TA^{(2)})=D_a T dx^a$.
Using trivial Chan--Paton factors, such that $F^{(i)}_{ab}=\prt_{a}A^{(i)}_{b}-\prt_{b}A^{(i)}_{a}$ and $D_{a}T=\prt_{a}T-i(A^{(1)}_a-A^{(2)}_a)T$, and following \cite{Kraus:2000nj}, the kinetic terms of the EFT action can be assembled into a DBI-like Lagrangian, viz.

\beqa
S_{DBI}&=&-\int
d^{p+1}\sigma {\rm TrS} \left(V({\cal T})
\sqrt{-\det(\eta_{ab}
+2\pi\alpha'F_{ab}+2\pi\alpha'D_a{\cal T}D_b{\cal T})} \right)\,\,,\labell{nonab} \eeqa 
where $V$ is the Tachyon potential, 
\beqa
F_{ab}:=\pmatrix{F^{(1)}_{ab}&0\cr 
0&F^{(2)}_{ab}}\ ,\qquad
D_{a}{\cal T}:=\pmatrix{0&D_aT\cr 
(D_aT)^*&0}\ ,\qquad {\cal T}=\pmatrix{0&T\cr 
T^*&0}\ ,\labell{M12} \eeqa 
and ${\rm TrS}$ denotes the symmetrized trace over ${\rm mat}_2$.
The EFT action also contains couplings to the spacetime RR potentials packaged into $C:=\sum_n(-i)^{\frac{p-n+1}{2}}C_n$ which can be assembled into a Wess--Zumino term 
\beqa
S_{WZ}&=&\mu_p \int_{\Sigma_{(p+1)}} C \wedge {\rm STr}\, e^{2\pi\alpha'i\cal F}\ ,\labell{WZ}\eeqa 
where ${\rm STr}$ is the supertrace.
Indeed, setting the tachyon modes to zero, the WZ term reduces to the Chern--Simons-like term \cite{Douglas:1995bn}\footnote{The nomenclature stems from the fact that $S_{\rm CS}$ consists of couplings of the form $\int C_n\wedge dA \wedge \cdots \wedge dA$, though since $C_n$ comprises the pull-back $\frac1{n!}C_{M_1\dots M_n} dX^{M_1}\wedge \cdots dX^{M_n}$, one may also think of $S_{\rm CS}$ as packaging worldvolume Quillen--Chern classes of a superconnection with components $(A,X^M,..)$ whose extension by $T$ thus deforms $S_{\rm CS}$ to $S_{\rm WZ}$. To observe  the exact forms of  Chern classes and superconnection relations \cite{quil,berl,Roepstorff:1998vh,Kraus:2000nj}  are recommended.}

\beqa
S_{\rm CS}:=S_{\rm WZ}|_{T=0} = \mu_p\int_{\Sigma_{(p+1)}} C \wedge  \left(e^{2\pi\alpha'iF^{(1)}}-e^{2\pi\alpha'iF^{(2)}}\right)\ ,
\labell{eqn.wz}
\eeqa
as expected by taking the naive sum of a separately supersymmetric $D$- and $\overline{\rm D}$-brane.

\paragraph{Comparison to disc amplitudes.} Expanding the Lagrangian of \reef{WZ}, one derives
\beqa
C\wedge \STr i{\cal F}&\!\!\!\!=\!\!\!&C_{p-1}\wedge(F^{(1)}-F^{(2)})\labell{exp2}\\
C\wedge \STr i{\cal F}\wedge i{\cal F}&\!\!\!\!=\!\!\!\!&C_{p-3}\wedge \left\{F^{(1)}\wedge F^{(1)}-
F^{(2)}\wedge F^{(2)}\right\}\nonumber\\
&& +C_{p-1}\wedge\left\{-2\beta^2|T|^2(F^{(1)}-F^{(2)})+2i\beta^2 DT\wedge(DT)^*\right\}\\
\nonumber\eeqa
the resulting couplings can be matched to disc amplitudes computed using CFT methods \cite{Friedan:1985ge}; for details, see for example \cite{Hatefi:2013yxa}.
To this end, the relevant open string vertex operators for real tachyons and gauge photons carrying Chan--Paton factors $\lambda$ are given by
\beqa
V_{T}^{(-1)}(x_1) &=&  e^{-\phi(x_1)}e^{\alpha'ik_1\cd X(x_1)}\lam\otimes\sigma_2 \ ,\\
V_{T}^{(0)}(x_2) &=&\alpha'ik_2\cd\psi(x_2) e^{\alpha'ik_2\cd X(x_2)}\lam\otimes\sigma_1\ , \\
V_A^{(-1)}(x_3)&=&{\xi}\cd {\psi(x_{3})}e^{-\phi(x_3)}\,e^{\alpha'ik_{3}\cd X(x_{3})}\labell{vertex2}\lam\otimes\sigma_3\ ,\\
V_A^{(0)}(x_4)&=&{\xi_{2a}}{\bigg(\partial X^a(x_4)+\alpha'ik\cd \psi(x_4) \psi^a(x_{4})\bigg)}e^{\alpha'ik_{4}\cd X(x_{4})}\lam\otimes I\ ,\eeqa
where $x_i$ belong to the boundary of the disc and $k_i$ are worldvolume momenta obeying $k_3^2=k_4^2=0$ and  $k_{1}^2=k_{2}^2=-m^2$, where $m^2=\frac{-1}{2\alpha'}$ is the tachyon mass-squared.
The closed string vertex operator for the RR particle is given by
\beqa
V_{RR}^{(-1)}(z,\bar{z})&=&(P_{-}\fsH_{n}M_p)^{\al\be}e^{-\phi(z)/2} S_{\al}(z)e^{i\frac{\alpha'}{2}p\cd X(z)}
e^{-\phi(\bar{z})/2} S_{\be}(\bar{z}) e^{i\frac{\alpha'}{2}p\cd D \cd X(\bar{z})}\otimes \sigma_3\ ,\quad\label{vertices}\eeqa
where $(z,\bar z)$ is a point inside the disc and $p$ a massless spacetime momentum, and\footnote{We raise and lower spinor indices using a charge conjugation matrix $C^{\alpha\beta}$ and the conventions $S^\alpha=C^{\alpha\beta}S_\beta$ and $S_\alpha=S^\beta (C^{-1})_{\beta\alpha}$.}
 \beqa
P_{-} &:=&\ha (1-\ga^{11})\ ,\\
\fsH_{n} &:=& \frac{m
_n}{n!}H_{\mu_{1}\ldots\mu_{n}}\ga^{\mu_{1}}\ldots
\ga^{\mu_{n}}\ ,
\eeqa
with $n$ even and $m_n=i$ in the Type IIA model, and $n$ odd and $m_n=1$ in the Type IIB model.
Using the doubling trick, the disc amplitude is mapped by 
\beqa
\tilde{X}^{\mu}(\bar{z}) \rightarrow D^{\mu}_{\nu}X^{\nu}(\bar{z}) \ ,
\spa
\tilde{\psi}^{\mu}(\bar{z}) \rightarrow
D^{\mu}_{\nu}\psi^{\nu}(\bar{z}) \ ,
\spa
\tilde{\phi}(\bar{z}) \rightarrow \phi(\bar{z})\ , \spa
\tilde{S}_{\al}(\bar{z}) \rightarrow M_{\al}{}^{\be}{S}_{\be}(\bar{z})
 \ ,
\eeqa
where the matrices
\beqa
D := \left( \begin{array}{cc}
-1_{9-p} & 0 \\
0 & 1_{p+1}
\end{array}
\right) \ ,\qquad
M_p := \frac{\pm m_p}{(p+1)!}\ga^{a_{0}}\ldots \ga^{a_{p}} (\gamma_{11})^p
\eps_{a_{0}\ldots a_{p}}\ ,
\eeqa
to an amplitude computed on the entire complex plane for a single holomorphic sector with two-point functions 
\begin{eqnarray}
\lan X^{\mu}(z)X^{\nu}(w)\ran & = & -\frac{\alpha'}{2}\eta^{\mu\nu}\log(z-w) \ ,  \\
\lan \psi^{\mu}(z)\psi^{\nu}(w) \ran & = & -\frac{\alpha'}{2}\eta^{\mu\nu}(z-w)^{-1} \ , \\
\lan\phi(z)\phi(w)\ran & = & -\log(z-w) \ .
\labell{prop2}\end{eqnarray}
Letting $u := -\frac{\alpha'}{2}(k_1+k_2)^2$ and setting $\alpha'=2$, the complete disc amplitude for the scattering of two real tachyons and one RR particle on the $D_p\bD_p$-brane reads 
 \beqa
{\cal A}^{T^{(-1)}T^{(0)}C^{(-1)}_{n-1}}_{\rm disc} &=&  \frac{i\mu_p}{4} 2\pi \frac{\Gamma(-2u)}{\Gamma(1/2-u)^2}\Tr(P_{-}\fsH_{n}M_p\gamma^a)k_{2a}\ ,
 \label{yy1}\eeqa
where $ \mu_p $ is RR charge and $\gamma^{11}$ is chosen such that $H_n=*H_{10-n}$ for  $p>3$ and $n\geq 5$.

In the limit $u=-p^ap_a\rightarrow 0$, the u-channel massless gauge field pole is reproduced in the EFT as 
\beqa
{\cal A}^{TTC_{p-1}}_{\rm EFT}|_{\rm u-ch}&=&V^a(C_{p-1},A^{(1)})G_{ab}(A)V^b(A^{(1)},T_1,T_2)\nonumber \\&&+V^a(C_{p-1},A^{(2)})G_{ab}(A)V^b(A^{(2)},T_1,T_2)\ ,\eeqa
where $G_{ab}(A)$ is the gauge field propagator, and $V_a(C_{p-1},A^{(i)})$ and $V_a(A^{(i)},T_1,T_2)$ are vertices read off, respectively, from the Chern-Simons coupling
\beqa
  i\mu_p (2\pi\alpha')\int_{\Sigma_{p+1}} \epsilon^{a_0...a_{p}}\Tr(C_{a_0...a_{p-2}}\partial_{a_{p-1}}(A^{(1)}_{a_{p}}-A^{(2)}_{a_{p}}))\ ,
\label{jj12}\eeqa
and the DBI-term.
Consequently, by matching the full EFT-amplitude to the disc amplitude, the required contact terms are contained in 
\beqa
i\mu_p (2\pi\alpha')^2 \int_{\Sigma_{p+1}} C_{p-1}\wedge \Tr\left(\sum_{m=-1}^{\infty}c_m(\alpha' (D ^a D_a))^{m+1}  DT_1 \wedge DT_2\right)\ , \labell{highaaw3}\eeqa
which thus provide the desired EFT couplings of two real tachyons and one RR potential to all orders in $\alpha'$. 

In what follows, we shall include one gauge photon into the amplitude and exhibit the resulting all-order bosonic four-field corrections to the $D_p\bD_p$-brane EFT.

\section{From S-matrix to All-Order EFT Couplings}\label{sec:higher}

In this section, we contextualize our analysis by reviewing the five point function, which also serves as a warm up for the computations of the new six point functions in the following section.

At tree level, the string scattering S-matrix element on a $D_p\bD_p$-brane of two real tachyons with momenta $k_2$ and $k_3$, a gauge photon with momentum $k_1$, and an RR particle with spacetime momentum $p$  is given by the disc amplitude \cite{Garousi:2007fk} 
\beqa {\cal A}^{A^{(-1)}T^{(0)}T^{(0)}C^{(-1)}_{n-1}}_{\rm disc}&=& \frac{i\mu_p}{2\sqrt{\pi}}\int dz d\bar z\,k_{2b}k_{3c}{\xi_a}(z-\bar z)^{-2(t+s+u)-1}|z|^{2t+2s-1}|1-z|^{2u+2t-1/2}\nonumber\\&&\times \bigg\{(\Gamma^{cba})_{\al\be}-2\h^{ab}\frac{-x+|z|^2}{z-\bar z}(\Gamma^{c})_{\al\be}-2\h^{ac}\frac{x}{z-\bar z}(\Gamma^{b})_{\al\be}\nonumber\\&&+2\h^{bc}\frac{x-1}{z-\bar z}(\Gamma^{a})_{\al\be}\bigg\}(P_{-}\fsH_{n}M_p)^{\al\be}\ ,\eeqa 
where we have gauge fixed $(x_1,x_2,x_3,z,\bar z)=(0,1,\infty,z,\bar{z})$, and introduced the Mandelstam variables
 \beqa
 s&:=&-\frac{\alpha'}{2}(k_1+k_3)^2,\quad t:=-\frac{\alpha'}{2}(k_1+k_2)^2,\quad u:=-\frac{\alpha'}{2}(k_2+k_3)^2\ .\eeqa 

Note that for five point functions the integration over the position of the RR potential vertex operator involves the prototype integral
\beqa
 \int  \int d^2 \!z |1-z|^{a} |z|^{b} (z - \bar{z})^{c}
(z + \bar{z})^{d}\ .
\eeqa
These integrals were computed for $d=0,1$ in \cite{Fotopoulos:2001pt} and for $d=2$ in \cite{Hatefi:2012wj} where the tricks are explained in \cite{Hatefi:2012wj}. Including a normalization $i\mu_p/2\sqrt{2\pi}$, the final form of the amplitude is
\beqa{\cal A}^{T^{(0)}T^{(0)}A^{(-1)}C^{(-1)}_{n-1}}_{\rm disc}&=&\frac{i\mu_p}{2\sqrt{2\pi}}\bigg(\Tr((P_{-}\fsH_{n}M_p)(k_3\cd\ga)(k_2\cd\ga)(\xi\cd\ga))I+\Tr((P_{-}\fsH_{n}M_p)\ga^{a})J\nonumber\\&&\times\bigg[k_{2a}(t+1/4)(2\xi\cd k_{3})+k_{3a}(s+1/4)(2\xi\cd k_{2})-\xi_a(s+1/4)(t+1/4)\bigg]\bigg)\ 
,\labell{gen}\nonumber\eeqa
where 
 \beqa I&:=&2^{-2(t+s+u)-1/2}\pi{\frac{\Gamma(-u) \Gamma(-s+1/4)\Gamma(-t+1/4)\Gamma(-t-s-u)}{\Gamma(-u-t+1/4)\Gamma(-t-s+1/2)\Gamma(-s-u+1/4)}}\ ,\\ J&:=&2^{-2(t+s+u)-3/2}\pi{\frac{\Gamma(-u+1/2)\Gamma(-s-1/4)\Gamma(-t-1/4)\Gamma(-t-s-u-1/2}{\Gamma(-u-t+1/4)\Gamma(-t-s+1/2)\Gamma(-s-u+1/4)}}\ .\eeqa 
Given the on-shell condition,  
\beqa
s+t+u&=&-p_ap^a-\frac{1}{2}\ .\label{ex1}\eeqa
The EFT formalism, which provide the guide of generating massless singularities, implies that $k_i.k_j \rightarrow 0$, which yields the following expansion for $CTTA$  amplitude:
\beqa
u\rightarrow 0\ ,\qquad s,t\rightarrow -1/4\ .\label{point}\eeqa

In \cite{Garousi:2007fk}, some tachyonic and massless singularities of the S-matrix are spelled out. 
For $p=n+2$, the amplitude has infinitely many massless poles, and for $p=n$ there are infinitely many tachyon and massless poles.
It does have an infinite number of gauge field poles in $(s+t+u+1/2)-$ channel poles as well. 

As for the infinite collection of massless gauge poles for $p=n+2$, the corresponding string amplitude is
\beqa
{\cal A}^{AT_1T_2C}&=&\pm\frac{8i\mu_p}{\sqrt{2\pi}(p-2)!} \eps^{a_{0}\cdots a_{p}}H_{a_{0}\cdots a_{p-3}}
k_{3a_{p-2}}k_{2a_{p-1}}\xi_{a_p} I\ ,\labell{pn2}\eeqa 
which are produced by the following sub amplitude in an EFT
\beqa
{\cal A}_{\rm EFT}&=&V_a(C_{p-3},A^{(1)},A^{(1)})G_{ab}(A^{(1)})V_b(A^{(1)},T_1,T_2)\ ,\labell{amp3}\eeqa
where the expansion of $I$ around \reef{point} is
\beqa
I&=&
\pi\sqrt{2\pi}\left(-\frac{1}{u}\sum_{n=-1}^{\infty}b_n(s'+t')^{n+1}+\sum_{p,n,m=0}^{\infty}c_{p,n,m}u^p\left(s't'\right)^n(s'+t')^m\right)\ ,\eeqa
where $s'=s+1/4, t'=t+1/4$; the algebraic structure of the coefficients is represented well by the first few ones, which are 
\beqa
&&c_{0,0,2}=\frac{2}{3}\pi^2\ln(2)\ ,\qquad c_{0,1,0}=-14\z(3)\ ,
\qquad c_{0,0,3}=8\z(3)\ln(2)\ .
\eeqa

Since the gauge field propagator has no correction, nor the couplings of two tachyons and a gauge field, it is necessary to add higher-derivative couplings between the RR potential two gauge fields, which can thus be determined to all orders, with the result
\beqa
\mu_p (2\pi\alpha')^2\sum_{n=-1}^{\infty}  b_n(\alpha')^{n+1} C_{p-3} \wedge D^{a_{1}}...D^{a_{n}} F\wedge D_{a_{1}}...D_{a_{n}}  F\ .
\eeqa
The infinite set massless poles can then be derived using the following data:
\beqa
G_{ab}(A^{(1)}) &=&\frac{i\delta_{ab}}{(2\pi\alpha')^2 T_p
\left(u\right)}\ ,\\
V_b(A^{(1)},T_1,T_2)&=&T_p(2\pi\alpha')(k_2-k_3)_b\\
V_a(C_{p-3},A^{(1)},A^{(1)})&=&\frac{\mu_p(2\pi\alpha')^2}{(p-2)!}\epsilon_{a_0\cdots a_{p-1}a}H^{a_0\cdots a_{p-3}}k_1^{a_{p-2}}\xi^{a_{p-1}}\sum_{n=-1}^{\infty}b_n(\alpha'k_1\cdot k)^{n+1}\ ,\eeqa
where $k$ is the momentum of the off-shell gauge field.
Indeed, it follows that all string poles are reproduced in the EFT, and the final form of the EFT amplitude \reef{amp3} is
\beqa
{\cal A}_{\rm EFT}=\mu_p(2\pi\alpha')\frac{2i}{(p-2)!u}\epsilon_{a_0\cdots a_{p-1}a}H^{a_0\cdots a_{p-3}}k_2^{a_{p-2}}k_3^{a_{p-1}}\xi^a\sum_{n=-1}^{\infty}b_n\left(\frac{\alpha'}{2}\right)^{n+1}(s'+t')^{n+1}\ .\labell{AA}\eeqa

On the other hand for $p=n$, one explores the following string amplitude:
\beqa
{\cal A}^{AT_1T_1C}&=&\pm\frac{8i\mu_p}{\sqrt{2\pi}p!} \eps^{a_{0}\cdots a_{p-1}a}H_{a_{0}\cdots a_{p-1}}J
\bigg\{k_{2a} t'(2\xi.k_{3})
+k_{3a}s'(2\xi.k_{2})-\xi_a s't'\bigg\}\ .\labell{pn22}\eeqa
Its first simple gauge field pole is reproduced by the following EFT amplitude:
    \beqa
{\cal A}_{EFT}&=&V_a(C_{p-1},A)G_{ab}(A)V_b(A,T_1,T_1,A^{(1)})\ ,\labell{amp4}\eeqa
where the vertices and gauge field propagators, which can be read off from the DBI action and CS term, are given by
   \beqa
G_{ab}(A) &=&\frac{i\delta_{ab}}{(2\pi\alpha')^2 T_p
\left(u+t'+s'\right)}\ ,\\
V_a(C_{p-1},A^{(1)})&=&i\mu_p(2\pi\alpha')\frac{1}{p!}\epsilon_{a_0\cdots a_{p-1}a}H^{a_0\cdots a_{p-1}}\labell{amp4'}\ ,\\
V_a(C_{p-1},A^{(2)})&=&-i\mu_p(2\pi\alpha')\frac{1}{p!}\epsilon_{a_0\cdots a_{p-1}a}H^{a_0\cdots a_{p-1}}\ ,\\
V_b(A^{1}, T_1,T_1,A^{1})&=& -2i T_p (2\pi \alpha') \xi_b\ ,\\
V_b(A^{(2)}, T_1, T_1, A^{(1)}) &=& 2i T_p (2\pi\alpha')\xi_b\ .\eeqa
 
On the other hand, tachyonic singularities are also produced by the following sub-amplitude: 
\beqa 
{\cal A}&=&V(C_{p-1},T_1,T_2)G(T_2)V(T_2,T_1,A^{(1)})\ ,\labell{finalA}
\eeqa
where we recall that the vertex of two tachyons and one gauge field has no correction as the kinetic term of tachyon is fixed and the propagator receives no correction either;  hence
\beqa
V(T_2,T_1,A^{(1)})&=&T_p(2\pi\alpha')(k_3-k)\inn\xi\ ,\\
G(T_2)&=&\frac{i}{(2\pi\alpha')T_p(s+1/4)}\ ,\labell{ver2}\\
V(C_{p-1},T_1,T_2)&=&(2\pi\alpha')^2\mu_p\sum_{n=0}^{\infty}b_n\frac{1}{p!} \eps^{a_{0}\cdots a_{p-1}a}H_{a_{0}\cdots a_{p-1}}(\alpha'k_1.k)^{n+1} k_{2a}\ ,\eeqa
where $k_{2a}$ is the momentum of the off shell tachyon. 

In order to derive all infinite massless poles , we need to obtain all-order $\alpha'$ corrections of two tachyons and two gauge fields in the worldvolume of brane-anti brane system and to do so we start to properly address a six point function including an RR, two tachyons and two gauge fields in the next section.

\section{EFT Couplings from Six Point Functions}\label{sec:higher}

Having warmed up with the five point function, we now proceed by first computing the S-matrix element for a closed-string RR state and two gauge fields and two tachyons on the worldvolume of the D$\overline{\rm D}$ system.
We then factorize it in a specific limit over gauge field poles that match to all orders in $\alpha'$ with corresponding higher-derivative couplings of the EFT of the D$\overline{\rm D}$ system.   

We start with the disk correlation functions   
\beqa
    (I_1^{cba})_{\alpha\beta}&=&\langle : S_\alpha(z) : S_\beta(\bar{z}) : \psi^a(x_2) : \psi^b(x_3) : \psi^c(x_4) : \rangle \, ,\\[5pt]
    (I_2^{cbaed})_{\alpha\beta} & =& \langle : S_\alpha(z) : S_\beta(\bar{z}) : \psi^d(x_1) \psi^e(x_1) : \psi^a(x_2) : \psi^b(x_3) : \psi^c(x_4) : \rangle \, ,
\eeqa
with the result
\beqa\label{eq:I1_expression}
(I_1^{cba})_{\alpha\beta} &=&\bigg\{(\Gamma^{cba})_{{\alpha\beta}}-\alpha' \eta^{ab}(\Gamma^{c})_{\alpha\beta} \, \frac{\Re(x_{25}x_{36})}{x_{23}x_{56}} +\alpha' \eta^{ac}(\Gamma^{b})_{\alpha\beta} \, \frac{\Re(x_{25}x_{46})}{x_{24}x_{56}}
\nonumber\\&&-\alpha' \eta^{bc}(\Gamma^{a})_{\alpha\beta} \, \frac{\Re(x_{35}x_{46})}{x_{34}x_{56}}\bigg)\bigg\} \, 2^{-\frac{3}{2}} \, x_{56}^{\frac{1}{4}} \, (x_{25}x_{26}x_{35}x_{36}x_{45}x_{46})^{-\frac{1}{2}} \, ,
\eeqa
where $x_5 \equiv z = x + i y \, , x_6 \equiv \bar z$ and $x_{ij} = x_i - x_j$, and 
\beqa\label{eq:I2_expression}
(I_2^{cbaed})_{\alpha\beta} &= 2^{-\frac{5}{2}} \, x_{56}^{\frac{5}{4}} \, (x_{25}x_{26}x_{35}x_{36}x_{45}x_{46})^{-\frac{1}{2}} \, 
(x_{15}x_{16})^{-1} \nonumber\\
&\times\bigg\{(\Gamma^{cbaed})_{{\alpha\beta}} \nonumber\\
& \quad{} - \frac{1}{2} \, \alpha' \, l_1 \, \frac{\Re(x_{15}x_{26})}{x_{12}x_{56}} - \frac{1}{2} \,
\alpha' \, l_2 \, \frac{\Re(x_{15}x_{36})}{x_{13}x_{56}} + \frac{1}{2} \, \alpha' \, l_3  \frac{\Re(x_{15}x_{46})}{x_{14}x_{56}}\nonumber\\
& \quad{} - \frac{1}{2} \, \alpha' \, l_4 \, \frac{\Re(x_{25}x_{36})}{x_{23}x_{56}} + \frac{1}{2} \,
\alpha' \, l_5 \, \frac{\Re(x_{25}x_{46})}{x_{24}x_{56}} - \frac{1}{2} \, \alpha' \, l_6 \, \frac{\Re(x_{35}x_{46})}{x_{34}x_{56}}
\nonumber\\
&
- \frac{1}{4} \, (\alpha')^2 \, l_7\bigg(\frac{\Re(x_{15}x_{26})}{x_{12}x_{56}}\bigg)\bigg(\frac{\Re(x_{35}x_{46})}{x_{34}x_{56}}\bigg) \nonumber\\
&
+ \frac{1}{4} \, (\alpha')^2 \, l_8
\bigg(\frac{\Re(x_{15}x_{36})}{x_{13}x_{56}}\bigg)\bigg(\frac{\Re(x_{25}x_{46})}{x_{24}x_{56}}\bigg)
\nonumber\\
&
- \frac{1}{4} \, (\alpha')^2 \, l_9 \bigg(\frac{\Re(x_{15}x_{46})}{x_{14}x_{56}}\bigg)\bigg(\frac{\Re(x_{25}x_{36})}{x_{23}x_{56}}\bigg)
\nonumber\\
&
+\frac{1}{4} \, (\alpha')^2 \, l_{10} \bigg(\frac{\Re(x_{15}x_{26})}{x_{12}x_{56}}\bigg)\bigg(\frac{\Re(x_{15}x_{36})}{x_{13}x_{56}}\bigg)
\nonumber\\
&
+\frac{1}{4} \, (\alpha')^2 \, l_{11} \bigg(\frac{\Re(x_{15}x_{26})}{x_{12}x_{56}}\bigg)\bigg(\frac{\Re(x_{15}x_{46})}{x_{14}x_{56}}\bigg)
\nonumber\\
&
+\frac{1}{4} \, (\alpha')^2 \, l_{12} \bigg(\frac{\Re(x_{15}x_{26})}{x_{12}x_{56}}\bigg)\bigg(\frac{\Re(x_{35}x_{46})}{x_{34}x_{56}}\bigg)
\bigg\}\ ,
\eeqa
where indices have been suppressed and we have defined the following tensor-bispinors:
\beqa\label{eq:l_structures}
l_1  &=& - 2 \eta^{ad}\Gamma^{cbe}
+ 2 \eta^{ea}\Gamma^{cbd}
\ ,\qquad l_2  = 2  \eta^{db}\Gamma^{cae} 
-2  \eta^{eb}\Gamma^{cad} \ ,\\[5pt]
 l_3  &=& 2  \eta^{dc}\Gamma^{bae}
 -2  \eta^{ec}\Gamma^{bad}\ ,\qquad
l_4  = 2  \eta^{ab}\Gamma^{ced}
\ ,\\[5pt] l_5  &=& 2  \eta^{ac}\Gamma^{bed} \ ,\qquad
l_6  = 2 \, \eta^{bc}\Gamma^{aed}\ ,\\[5pt]
l_7  &=& 4 \eta^{ad}\eta^{bc}\Gamma^{e}
-4 \eta^{ea}\eta^{bc}\Gamma^{d}\ ,\\[5pt]
l_8 &=& 4 \, \eta^{db}\eta^{ac}\Gamma^{e}
-4 \, \eta^{eb}\eta^{ac}\Gamma^{d}\ ,\\[5pt]
l_9 & =& 4  \eta^{dc}\eta^{ab}\Gamma^{e} 
-4  \eta^{ec}\eta^{ab}\Gamma^{d}\ ,\\[5pt]
l_{10}  &=& 4  \eta^{da}\eta^{eb}\Gamma^{c} 
+4  \eta^{ea}\eta^{db}\Gamma^{c}\ ,\\[5pt]
l_{11}  &=& 4  \eta^{da}\eta^{ec}\Gamma^{b} 
-4  \eta^{ea}\eta^{dc}\Gamma^{b}\ ,\\[5pt]
l_{12}  &=& -4  \eta^{da}\eta^{bc}\Gamma^{d}
+4  \eta^{ea}\eta^{bc}\Gamma^{d}\ .
\eeqa
The desired S-matrix element can then be computed, with the result
\begin{eqnarray}\label{eq:amplitude1}
  A^{C^{(-1)} A^{(0)} A^{(-1)} T^{(0)} T^{(0)}}&=& 2i  \Tr(\lambda_1 \lambda_2 \lambda_3 \lambda_4)  (P_{-}\fsH_{n}M_p)^{\alpha \beta} \, \xi_{1e} \, \xi_{2a}\nonumber\\
& &\quad{}\times\int dx_1 dx_2
dx_3 dx_4 dx_5 dx_6 \, I  \, x_{56}^{-\frac{1}{4}} \, (x_{25}x_{26})^{-\frac{1}{2}}\nonumber\\
& &\quad{}\times\bigg( i(\alpha')^2 \, k_{3b} k_{4c} \bigg[p^e(  \frac{x_{25}}{x_{15}x_{12}}+\frac{x_{26}}{x_{16}x_{12}})+2k^e_{3}\frac{x_{23}}{x_{13}x_{12}}+2k^e_{4}\frac{x_{24}}{x_{14}x_{12}}\bigg] I_1^{cba}\nonumber\\
& &- i \, (\alpha')^3  k_{1e}  k_{3b} k_{4c} \, I_2^{cbaed} \bigg)\ ,
\end{eqnarray}
where
\beqa\label{eq:I_factor}
I & = |x_{12}|^{-2t}|x_{13}|^{-2s-\frac{1}{2}}|x_{14}|^{-2v-\frac{1}{2}}|x_{23}|^{-2u-\frac{1}{2}}
|x_{24}|^{-2r-\frac{1}{2}}|x_{34}|^{-2w-1}|x_{15}x_{16}|^{t+s+v+\frac{1}{2}} \nonumber\\
& \quad{} \times
|x_{25}x_{26}|^{t+u+r+\frac{1}{2}}|x_{35}x_{36}|^{s+u+w+\frac{1}{2}}|x_{45}x_{46}|^{v+r+w+\frac{1}{2}}|x_{56}|^{-2(s+t+u+v+r+w)-2} \, ,
\eeqa
in terms of the six independent Mandelstam variables
\beqa
    s &=& -(\frac{1}{4} + 2 k_1 \cdot k_3)\ ,\qquad 
    t = - 2 k_1 \cdot k_2\ ,\qquad  v = - ( \frac{1}{4} + 2 k_1 \cdot k_4 )\ ,\\
    u &=& -( \frac{1}{4} + 2  k_2 \cdot k_3 )\ ,\qquad 
    r  = -( \frac{1}{4} + 2 k_2 \cdot k_4)\ ,\qquad  
    w = -( \frac{1}{2} + 2 k_3 \cdot  k_4 )\ .
\eeqa
Since the amplitude has been written so that it respects $SL(2,R)$ invariance manifestly, we can fix the positions of three open-string vertex operator; we choose
\begin{equation}
    x_1 = 0 \, , \qquad  0 \leq x_2 \leq 1 \, , \qquad  x_3 = 1 \, , 
\qquad x_4 = \infty \,,
\end{equation}
which brings the final amplitude to the form
\begin{eqnarray}
&&-8k_{3b}k_{4c}\xi_{1e}\xi_{2a}  2^{-\frac{1}{2}} (P_{-}\fsH_{n}M_p)^{\alpha\beta}\int_0^1 dx_2 x_2^{-2t} (1-x_2)^{-2u-\frac{1}{2}}
\nonumber\\
&&\times\int d^2 z |1-z|^{2s+2u+2w} \, |z|^{2t+2s+2v-1} (z - \bar{z})^{-2(t+s+u+v+r+w)-2}  |x_2-z|^{2t+2u+2r-1}\nonumber\\
&&\times
 \bigg\{\bigg(p^e x^{-1}_{2}\frac{x_2(z+\bar z)-2|z|^2}{|z|^2}+2k^e_{3}(x_2-1)-2k^e_{4}\bigg)\\
 &&\times
   \bigg( (z-\bar z)(\Gamma^{cba})_{{\alpha\beta}} + 2  \eta^{ab} \, (\Gamma^{c})_{\alpha\beta}  \frac{x_2-x \, x_2-x+|z|^{2}}{(1-x_2)}
\nonumber\\
 && + 2  \eta^{ac}(\Gamma^{b})_{\alpha\beta}(x-x_2)+2 \, \eta^{bc} \, (\Gamma^{a})_{\alpha\beta}(1-x)\nonumber    \bigg)     \\
&&-2k_{1d}  \bigg[ (z -\bar z)
(\Gamma^{cbaed})_{\alpha\beta}+l_1 \, \frac{-x \, x_2+|z|^{2}}{x_2} + l_2 \left(- x+|z|^{2} \right) + l_3 \, x\nonumber\\
&&
+ l_4 \, \frac{x_2-x \, x_2-x+|z|^{2}}{(1-x_2)} + l_5 \left(x-x_2\right) + l_6 \left(1-x\right) \nonumber\\
&&
+ (z - \bar{z})^{-1}\bigg( l_7 \, (-1+x) \, \frac{-x \, x_2+|z|^{2}}{x_2} + l_8 \left(-x+|z|^{2} \right) (x_2-x) 
\nonumber\\
&&
+ l_9  x  \frac{x_2-x  x_2-x+|z|^{2}}{(1-x_2)} \bigg)
+l_{10} x  \frac{-x+|z|^{2}}{(-1+x_2)}\nonumber\\
&&
+l_{11} x  \frac{x-|z|^{2}}{(x_2)}
+l_{12} (-x+|z|^{2}+x^2-x|z|^2)
\bigg]\bigg\}\Tr(\lambda_1 \lambda_2 \lambda_3 \lambda_4) \, .
\end{eqnarray}
The total amplitude can be expressed in terms of the integrals 
\beqa\label{eq:total_amplitude_def}
    A_{xz}(a,b,c,d\mid \alpha,\beta\mid\epsilon) & \equiv \int d^2z \int_0^1 dx\, |{1-z}|^a |{z}|^b (z-\bar{z})^c (z+\bar{z})^d   \nonumber\\
    & \times (1-x)^\alpha x^\beta\, |{x - z}|^\epsilon,
\eeqa
whose properties were described in detail in \cite{Antonelli:2019pzg}.  
As explained in \cite{Antonelli:2019pzg}, one can work with the soft limit of the six point function as $4k_2\cdot p\rightarrow 1$ which gives an analytic expression for the amplitude, and some of the integrals can be found in Appendix B of  \cite{Fotopoulos:2001pt}.
They are subject to the on-shell condition 
\begin{equation}\label{eq:mandelstam_sum}
    s+t+u+v+r+w = - \, p^a p_a -1 \, ,
\end{equation}

In what follows, we shall determine quartic couplings in the EFT action by subtracting EFT exchange graphs with massless poles from the $TTAAC$ disc amplitude in the limit where $k_i\cd k_j \rightarrow 0$ for $i\neq j$.
To investigate the structure of poles, we  consider the correct limit of the Mandelstam variables to set up the low-energy expansion where the correct expansion is 
\beqa
\label{eq:limittsvpw}
t \rightarrow  0 \, , \quad s \, , v \, , - \, p^a p_a  \rightarrow  - \frac{1}{4} \,,\quad  w \rightarrow  - \frac{1}{2} \, ,\nonumber\\
\frac{1}{2}\left[(u\rightarrow 0 \, , r \rightarrow -\frac{1}{4}) + (u\rightarrow -\frac{1}{4} \, , r
\rightarrow 0) \right] \, .
\eeqa
where in the last line  we took the averaged over the two limits and  \reef{eq:limittsvpw} is consistent with momentum conservation,
and conforms to the symmetries of the S-matrix. Note that in the limit $-p^a p_a \rightarrow -\frac{1}{4}$, we conform to the RR-tachyon two-point function, for which $p^a p_a = \frac{1}{4}$ is a constraint.

Next, we derive the precise form of the all-order corrections to the couplings of two tachyons and two gauge fields with exact coefficients in the worldvolume of the D$\overline{\rm D}$ system.
We will then show that they reproduce the all-order gauge field poles in an EFT analysis.
By algebraic computations, we will also show that the corresponding existing proposal in the literature is inconsistent.

The string amplitude is non-vanishing in the case of RR potentials given by $C_{p-5}$, $C_{p-3}$ and $C_{p-1}$.
In the $C_{p-5}$ case, the amplitude has just contact terms as follows:
\beqa\label{eq:A1}
    A_{1} = \frac{64i 2^{-\frac{1}{2}}  \pi   }{(p-4)!}  N_1  N_2 k_{4c} k_{3b}\xi_{1e}  \xi_{2a}k_{1d} \, \epsilon^{a_0\ldots a_{p-5}cbaed} H_{a_0 \ldots a_{p-5}}\ ,  \eeqa
where 
\beqa
    N_1  & =& 2^{-2(t+s+u+v+r+w)-1}  \frac{\Gamma(-2u+\frac{1}{2}) \ \Gamma(-2t+1)}{\Gamma(-2t-2u+\frac{3}{2})}\ , \eeqa
    \beqa
    N_2  & =& \frac{\Gamma(-u-r-w) \, \Gamma(-t-v-r+\frac{1}{2}) \, \Gamma(r-s) \, \Gamma(-t-s-u-v-r-w)}{\Gamma(-u-s-w) \, \Gamma(-t-s-v+\frac{1}{2}) \, \Gamma(-u-w-t-v-2r+\frac{1}{2})} \ . 
\eeqa
It is readily seen that the leading contact terms can be reproduced by the EFT couplings
\beqa
\mu_p (2\pi\alpha')^4\int \sum_{n=-1}^{\infty}  b_n(\alpha')^{n+1} C_{p-5} \wedge F
\wedge F
\wedge DT_1
\wedge DT_2\ .
\eeqa
Other contact terms can be reproduced by properly adding higher derivative corrections; for more details see \cite{Antonelli:2019pzg}.
The amplitude has infinitely many massless poles in the $C_{p-3}$ case, and there are infinitely many tachyon poles in the $C_{p-1}$ case, where, for the sake of obtaining correct higher derivative corrections, it is sufficient to reconstruct the infinite set of gauge field poles.

For $p=n+2$, the amplitude is given by 
\beqa
{\cal A}^{AATTC_{p-1}}&=&\frac{8i\mu_p}{\sqrt{2\pi}(p-2)!} \epsilon^{a_0\cdots a_{p-1}a}H_{a_0\cdots a_{p-3}}k_{2a_{p-2}}\xi_{a_{p-1}} \nonumber\\&&\times
\bigg(k_{3a}u'(2\xi.k_{4})+k_{4a}r'(2\xi.k_{3})
-\xi_a u' r'\bigg) N_3 N_4
\ ,\labell{pn}\eeqa
where $N_3, N_4$ are the multiplications of various Gamma functions; for their explicit forms, see Appendix B of \cite{Antonelli:2019pzg}.

Having taken the proper expansion, one can show that all massless poles in the string amplitude are given by
\beqa
\frac{8i\mu_p}{\sqrt{2\pi}(p-2)!}\frac{1}{(w'+r'+u')} \epsilon^{a_0\cdots a_{p-1}a}H_{a_0\cdots a_{p-3}}k_{2a_{p-2}}\xi_{a_{p-1}}\nonumber\\
\bigg(k_{3a}u'(2\xi.k_{4})+k_{4a}r'(2\xi.k_{3})
-\xi_a u' r'\bigg)  \sum_{n,m=0}^{\infty}d_{n,m}(u'+r')^n(u'r')^m\ ,\labell{masslesspole}\eeqa
using momentum conservation in directions parallel to the branes, $(k_1 + k_2 + k_3 + k_4 + p)^a=0$, and the Bianchi identity 
\begin{equation}\label{eq:bianchi2}
      p_{b} \, H_{a_0\cdots a_{p-4}} \epsilon^{a_0\cdots a_{p-1}b}=0\,.
\end{equation}

\section{All-Order Two-Tachyon-Two-Gauge Field Couplings}

In this section, we determine the all-order part of the EFT Lagrangian on the worldvolume of a D$\overline{\rm D}$ system that is separately quadratic in gauge fields and tachyons, following the procedure used in \cite{Hatefi:2017ags}\footnote{The S-matrix element for scattering of one RR particle and four tachyons on a D$\overline{\rm D}$-brane is given in \cite{Hatefi:2017ags}, and its relation to the Veneziano amplitude \cite{Veneziano:1968yb} is spelled out in there as well.
Moreover, the all-order corrections to fermionic couplings  \cite{Hatefi:2019rxq} as well as a power series representation of the corresponding string amplitudes were obtained well beyond the factorized limit in \cite{Antonelli:2019pzg} .} 
to compute the highly-symmetric four-tachyon vertices to all orders in $\alpha'$ in the non-BPS case; we shall then verify the result by comparing the resulting gauge field poles against those of the corresponding string amplitude.
Our results modify and complete previous simpler proposals \cite{Garousi:2007fk}, whose subtle shortcomings we shall thus begin by highlighting before turning to the main part of this section.

\paragraph{Old proposal.} The vertices $V_a(A,A^{(1)},T_1,T_2)$ stem from the part of the EFT Lagrangian of the D$\overline{\rm D}$ system that is separately quadratic in gauge fields and tachyons.
In what follows, we will first review the tachyon-two gauge field EFT couplings of non-BPS branes that have been proposed in the literature \cite{Garousi:2007fk}, and show that these do not reproduce the desired string disc amplitude in the case of the D$\overline{\rm D}$ system, after which we shall provide a remedy in the form of a modified set of EFT higher-derivative couplings for this system. 
To spell out the couplings, it is useful to define the quadri-linear higher-derivative operators\footnote{If $D$ is a differential operator acting in an algebra of functions, then we use the convention $Dfg\equiv (Df)g$.} 
\beqa
D_{nm}(E,F,G,H)&:=&D_{b_1}\cdots D_{b_m}D_{a_1}\cdots D_{a_n} E F D^{a_1}\cdots D^{a_n}G D^{b_1}\cdots D^{b_m}H\ ,\\[5pt]
D'_{nm}(E,F,G,H)&:=&D_{b_1}\cdots D_{b_m}D_{a_1}\cdots D_{a_n}E   D^{a_1}\cdots D^{a_n}F G D^{b_1}\cdots D^{b_m}H\ .\eeqa
The existing proposal for the EFT Lagrangian of D$\overline{\rm D}$ systems consists of the leading couplings
 \beqa T_p(\pi\alpha')^3{\rm
TrS} \left(
\frac{}{}m^2{\cal T}^2F_{\mu\nu}F^{\nu\mu}+\frac{}{}D^{\alpha}{\cal T} D_{\alpha}{\cal T} F_{\mu\nu}F^{\nu\mu}-
4F^{\mu\alpha}F_{\alpha\beta}D^{\beta}{\cal T} D_{\mu}{\cal T}
\right)\labell{dbicoupling}\ , \eeqa 
at order $(\alpha')^3$ and where TrS denotes the symmetrised trace, combined with the higher-order couplings
\beqa
\cL^{TTAA}_{\rm old\,EFT}&=&-T_p(\pi\alpha')\sum_{n,m=0}^{\infty}(\alpha')^{2+n+m}(\cL_{{\rm old}\,1}^{nm}+\cL_{{\rm old}\,2}^{nm}+\cL_{{\rm old}\,3}^{nm}+\cL^{nm}_{{\rm old}\,4})\ ,\labell{lagrang}\eeqa
where\footnote{The first few $a_{n,m}$ and $b_{n,m}$ coefficients are given by 
\beqa
&&a_{0,0}=-\frac{\pi^2}{6},\,b_{0,0}=-\frac{\pi^2}{12}\\
&&a_{1,0}=2\z(3),\,a_{0,1}=0,\,b_{0,1}=b_{1,0}=-\z(3)\nonumber\\
&&a_{1,1}=a_{0,2}=-7\pi^4/90,\,a_{2,0}=-4\pi^4/90,\,b_{1,1}=-\pi^4/180,\,b_{0,2}=b_{2,0}=-\pi^4/45\nonumber\\
&&a_{1,2}=a_{2,1}=8\z(5)+4\pi^2\z(3)/3,\,a_{0,3}=0,\,a_{3,0}=8\z(5),\nonumber\\
&&\qquad\qquad\qquad\qquad\qquad b_{0,3}=-4\z(5),\,b_{1,2}=-8\z(5)+2\pi^2\z(3)/3\nonumber\eeqa}
\beqa
\cL_{{\rm old}\,1}^{nm}&=&m^2{\rm TrS}\left(\frac{}{}a_{n,m}[D_{nm}({\cal T}, {\cal T} ,F_{\mu\nu}, {F}^{\nu\mu})+ D_{nm}(F_{\mu\nu} ,{F}^{\nu\mu}, {\cal T}, {\cal T})]\right.\nonumber\\
&&\left.+\frac{}{}b_{n,m}[D'_{nm}({\cal T} , F_{\mu\nu} ,{\cal T},  F^{\nu\mu})+D'_{nm}( F_{\mu\nu}, {\cal T} , F^{\nu\mu}, {\cal T})]+c.c.\right)\ ,\\
\cL_{{\rm old}\,2}^{nm}&=&{\rm TrS}\left(\frac{}{}a_{n,m}[D_{nm}(D^{\alpha}{\cal T},  D_{\alpha}{\cal T}, F_{\mu\nu}, {F}^{\nu\mu})+D_{nm}( F_{\mu\nu}, {F}^{\nu\mu} ,D^{\alpha}{\cal T}, D_{\alpha}{\cal T})]\right.\nonumber\\
&&\left.+\frac{}{}b_{n,m}[D'_{nm}(D^{\alpha} {\cal T} ,F_{\mu\nu} ,D_{\alpha} {\cal T}, F^{\nu\mu})+D'_{nm}( F_{\mu\nu} ,D_{\alpha} {\cal T}, F^{\nu\mu}, D^{\alpha} {\cal T})]+c.c.\right)\ ,\\
\cL_{{\rm old}\,3}^{nm}&=&-2{\rm TrS}\left(\frac{}{}a_{n,m}[D_{nm}(D^{\beta}{\cal T} ,D_{\mu}{\cal T} ,{F}^{\mu\alpha}, F_{\alpha\beta})+D_{nm}( {F}^{\mu\alpha}, F_{\alpha\beta} ,D^{\beta} {\cal T}, D_{\mu} {\cal T})]\right.\nonumber\\
&&\left.+\frac{}{}b_{n,m}[D'_{nm}(D^{\beta} {\cal T} ,{F}^{\mu\alpha} ,D_{\mu}{\cal T}, F_{\alpha\beta})+D'_{nm}({F}^{\mu\alpha}, D_{\mu}{\cal T} ,F_{\alpha\beta}, D^{\beta} {\cal T})]+c.c.\right)\ ,\quad\\
\cL_{{\rm old}\,4}^{nm}&=&-2{\rm TrS}\left(\frac{}{}a_{n,m}[D_{nm}(D^{\beta} {\cal T}, D_{\mu} {\cal T},  F_{\alpha\beta} ,{F}^{\mu\alpha})+D_{nm}( F_{\alpha\beta}, {F}^{\mu\alpha}, D^{\beta}{\cal T} ,D_{\mu} {\cal T})]\right.\nonumber\\
&&\left.+\frac{}{}b_{n,m}[D'_{nm}(D^{\beta}{\cal T} ,F_{\alpha\beta} ,D_{\mu} {\cal T}, {F}^{\mu\alpha})+D'_{nm}( F_{\alpha\beta}, D_{\mu} {\cal T}, {F}^{\mu\alpha},D^{\beta} {\cal T})]+c.c.\right)\ .
\eeqa
Using $\cL^{TTAA}_{\rm old}$, the EFT sub-amplitude  
\beqa
{\cal A}^{CAATT}_{\rm gauge-ch}&=&V^a(C_{p-3},A,A)G_{ab}(A)V^b(A,A,T,T)\labell{finalamp44}\eeqa
contains the following infinite set of gauge poles 
\beqa
8i\mu_p \eps^{a_{0}\cdots a_{p-1}a} H_{a_{0}\cdots a_{p-3}} k_{1a_{p-2}}
\xi_{1a_{p-1}}
 \frac{1}{(p-2)!(u'+r'+w')}
\sum_{n,m=0}^{\infty}(a_{n,m}+b_{n,m})\label{88}\\ (u'^mr'^n+u'^n r'^m) 
[k_{3a}u'(2\xi.k_{4})+k_{4a}r'(2\xi.k_{3})
-\xi_a u' r']\ .
\nonumber\eeqa
At zeroth order in $\alpha'$, the residue is given by
\beqa
4 u' r'(a_{0,0} + b_{0,0}) &=& -\pi^2 u' r'\ ,
\eeqa
while the corresponding quantity in the disc amplitude is given by
\beqa
d_{0,0}u'r'=\frac{-\pi^2}{3} u'r'=d_{0,0}u'r'\ .
\eeqa
Clearly, the old EFT proposal does not reproduce the string result. 
Indeed, this discrepancy prevails at higher orders in the $\alpha'$-expansion; for example, at the first order in $\alpha'$, the EFT residue is given by
 \beqa
2 u' r' (u'+r')(a_{1,0} + a_{0,1}+b_{1,0} + b_{0,1}) &=& 0\ ,
\eeqa
in disagreement with the string theory results, which reads
\beqa
d_{1,0}u'r' (u'+r')=8 \xi(3) u'r'(u'+r')\ .
\eeqa
In other words, the proposed EFT coulings do not apply to the D$\overline{\rm D}$ configuration at hand.

\paragraph{New proposal.} An improved EFT Lagrangian for the D$\overline{\rm D}$ system can be found using the method of \cite{Hatefi:2017ags}; the sought for all-order expression for the DBI--SYM Lagrangian is given by 
\beqa {\cal
L}_{DBI}&\!\!\!=\!\!\!&-T_p(2\pi\alpha')\left(m^2 |T|^2+DT\cdot(DT)^{*}-\frac{\pi\alpha'}{2}
\left(F^{(1)}\cdot{F^{(1)}}+
F^{(2)}\cdot{F^{(2)}}\right)\right)+T_p(\pi\alpha')^3\nonumber\\
&&\times\left(\frac{2}{3}DT\cdot(DT)^{*}\left(F^{(1)}\cdot{F^{(1)}}+F^{(1)}\cdot{F^{(2)}}+F^{(2)}\cdot{F^{(2)}}\right)\right.\nonumber\\
&&\left.+\frac{2m^2}{3}|T|^2\left(F^{(1)}\cdot{F^{(1)}}+F^{(1)}\cdot{F^{(2)}}+F^{(2)}\cdot{F^{(2)}}\right)\right.\nonumber\\
&&-\left.\frac{4}{3}\left((D^{\mu}T)^*D_{\beta}T+D^{\mu}T(D_{\beta}T)^*\right)\left({F^{(1)}}^{\mu\alpha}F^{(1)}_{\alpha\beta}+{F^{(1)}}^{\mu\alpha}F^{(2)}_{\alpha\beta}+{F^{(2)}}^{\mu\alpha}F^{(2)}_{\alpha\beta}\right)\right)\ ,
\labell{exp1}
\eeqa
at the leading order in the $\alpha'$-expansion, and
\beqa
\cL^{TTAA}_{\rm EFT}&=&-T_p(\pi\alpha')\sum_{n,m=0}^{\infty}(\alpha')^{2+n+m}(\cL_{1}^{nm}+\cL_{2}^{nm}+\cL_{3}^{nm}+\cL^{nm}_{4})\ ,\labell{lagrang}\eeqa
to higher orders, where
\beqa
\cL_1^{nm}&=&m^2
{\rm TrS}\left(\frac{}{}a_{n,m}[D_{nm}(\cT ,\cT^*, F^{(1)}_{\mu\nu} ,{F}^{(1)\nu\mu})+ D_{nm}(F^{(1)}_{\mu\nu} ,{F}^{(1)\nu\mu}, \cT ,\cT^*)]\right.\nonumber\\
&&\left.-\frac{}{}b_{n,m}[D'_{nm}(\cT  ,F^{(2)}_{\mu\nu} ,\cT^* , F^{(1)\nu\mu})+D'_{nm}( F^{(1)}_{\mu\nu} ,\cT , F^{(2)\nu\mu} ,\cT^*)]+c.c.\right)\ ,\\
\cL_2^{nm}&=&{\rm TrS}\left(\frac{}{}a_{n,m}[D_{nm}(D^{\alpha}\cT , D_{\alpha}\cT^*, F^{(1)}_{\mu\nu}, {F}^{(1)\nu\mu})+D_{nm}( F^{(1)}_{\mu\nu}, {F}^{(1)\nu\mu} D^{\alpha}\cT, D_{\alpha}\cT^*)]\right.\nonumber\\
&&\left.-\frac{}{}b_{n,m}[D'_{nm}(D^{\alpha} \cT ,F^{(2)}_{\mu\nu}, D_{\alpha}\cT^* ,F^{(1)\nu\mu})+D'_{nm}( F^{(1)}_{\mu\nu} ,D_{\alpha}\cT ,F^{(2)\nu\mu} ,D^{\alpha} \cT^*)]+c.c.\right)\ ,\\
\cL_3^{nm}&=&-2{\rm TrS}
\left(\frac{}{}a_{n,m}[D_{nm}(D^{\beta}\cT, D_{\mu}\cT^* ,{F}^{(1)\mu\alpha} ,F^{(1)}_{\alpha\beta})+D_{nm}( {F}^{(1)\mu\alpha}, F^{(1)}_{\alpha\beta}, D^{\beta}\cT ,D_{\mu}\cT^*)]\right.\nonumber\\
&&\left.-\frac{}{}b_{n,m}[D'_{nm}(D^{\beta}\cT ,{F}^{(2)\mu\alpha}, D_{\mu}\cT^*, F^{(1)}_{\alpha\beta})+D'_{nm}({F}^{(1)\mu\alpha}, D_{\mu}\cT, F^{(2)}_{\alpha\beta}, D^{\beta}\cT^*)]+c.c.\right)\ ,\\
\cL_4^{nm}&=&-2{\rm TrS}\left(\frac{}{}a_{n,m}[D_{nm}(D^{\beta}\cT ,D_{\mu}\cT^*,  F^{(1)}_{\alpha\beta}, {F}^{(1)\mu\alpha})
+D_{nm}( F^{(1)}_{\alpha\beta}, {F}^{(1)\mu\alpha}, D^{\beta}\cT, D_{\mu}\cT^*)]\right.\nonumber\\
&&\left.-\frac{}{}b_{n,m}[D'_{nm}(D^{\beta}\cT, F^{(2)}_{\alpha\beta}, D_{\mu}\cT^*, {F}^{(1)\mu\alpha})+D'_{nm}( F^{(1)}_{\alpha\beta}, D_{\mu}\cT, {F}^{(2)\mu\alpha},D^{\beta}\cT^*)]+c.c.
\right)\ ,\qquad\quad
\eeqa
where $b_{n,m}=b_{m,n}$, and $F^{(1)}$ and $F^{(2)}$, respectively, are the gauge field strengths on the D-brane and $\overline{D}$-brane. 
We note that unless the field strengths on the D- and $\overline{D}$-brane are treated separately, the EFT residues in the gauge channel cannot be made to match the string results even to the leading order in $\alpha'$.

The modification of the EFT Lagrangian proposed here can also be motivated by the structure of EFT couplings on the D$\overline{\rm D}$ system including fermions, viz. $\bar\psi^{2}\gamma^aT\psi^1D_aT$ and $\psi^1 D_aT\bar\psi ^{2}\gamma^a T$, found recently in \cite{Hatefi:2019rxq}.

\paragraph{Reproduction of gauge field poles.} Let us consider again the EFT sub-amplitudes with massless gauge fields poles, viz.
\beqa
{\cal A}^{CAATT}_{\rm EFT}&=&V_a(C_{p-3},A,A^{(1)})G_{ab}(A)V_b(A,A^{(1)},T_1,T_1)\ ,\labell{finalamp}\eeqa
where, working perturbatively, the vertex $V_a(C_{p-1},A)$ and propagator $G_{ab}(A)$ are  given, provided that one uses the following gauge propagator and vertices:
\beqa
G_{ab}(A) &=&\frac{i\delta_{ab}}{(2\pi\alpha')^2 T_p (u'+r'+w')}\ ,\\
V_a(C_{p-3},A,A^{(1)})&=&\frac{\mu_p(2\pi\alpha')^2}{(p-2)!}\epsilon_{a_0\cdots a_{p-1}a}H^{a_0\cdots a_{p-3}}k_1^{a_{p-2}}\xi^{a_{p-1}}\sum_{n=-1}^{\infty}b_n(\alpha'k_1\cdot k)^{n+1}\ ,\\
V_a(C_{p-3},A,A^{(2)})&=&-\frac{\mu_p(2\pi\alpha')^2}{(p-2)!}\epsilon_{a_0\cdots a_{p-1}a}H^{a_0\cdots a_{p-3}}k_2^{a_{p-2}}\xi^{a_{p-1}}\sum_{n=-1}^{\infty}b_n(\alpha'k_2\cdot k)^{n+1}\ ,\quad
\eeqa
where $k$ is the momentum of the off-shell gauge field.
Thus, the all-order form of the vertex $V_a(C_{p-3},A,A^{(1)})$ is determined from
\beqa
\mu_p (2\pi\alpha')^2\sum_{n=-1}^{\infty}  b_n(\alpha')^{n+1} C_{p-3} \wedge D^{a_{0}}...D^{a_{n}} F\wedge D_{a_{0}}...D_{a_{n}}  F\ .
\eeqa
To the leading order in the $\alpha'$-expansion, we take the vertex $V_a(A,A^{(1)},T_1,T_2)$ from the second line of \reef{exp1}, with the result
\beqa
V_a(A^{(1)},A^{(1)},T_1,T_1)&\!\!\!=\!\!\!&2iT_p(\pi\alpha')^3\left[\frac23 k_a\left[r'(2k_3\cdot\xi)+u'(2k_4\cdot\xi)\right]\right.\\
&&\left.\phantom{\frac23}+k_{3a}u'(2\xi.k_{4})+k_{4a}r'(2\xi.k_{3})
-\xi_a u' r'
\right]\ ,\nonumber\\
V_a(A^{(2)},A^{(1)},T_1,T_2)&\!\!\!=\!\!\!&2iT_p(\pi\alpha')^3\left[\frac13 k_a\left[r'(2k_3\cdot\xi)+u'(2k_4\cdot\xi)\right]+\right.\\
&&\left.\phantom{\frac13}
+k_{3a}u'(2\xi.k_{4})+k_{4a}r'(2\xi.k_{3})
-\xi_a u' r'
\right]\ ,\nonumber
\eeqa
where $k^a$ is the momentum of the off-shell gauge field. 
Indeed, substituting the above vertices into \reef{finalamp} reproduces the first massless singularity of \reef{masslesspole}.

Note that the correction to the vertex is read off by 
\beqa
-2\alpha'\mu_p \sum_{n=0}^{\infty}a_n\left(\frac{\alpha'}{2}\right)^n C_{p-1}\wedge (D^aD_a)^n[(F^{(1)}-F^{(2)})|T|^2]\ ,\labell{hderv2}
\eeqa
which is the extension of the couplings that are given in \reef{exp2}.

Proceeding to higher orders, we use the two-tachyon-two-gauge field couplings from \reef{lagrang}, which read
\beqa
&&4iT_p
(\alpha')^{n+m}(a_{n,m}-b_{n,m})\Tr(\lam_2\lam_3\lam_4\Lambda^{\beta})\bigg[\bigg((k_2\inn k_3)^m(k_2\inn k_4)^n+(k_2\inn k_3)^n(k_2\inn k_4)^m\nonumber\\&&
(k_1\inn k)^m(k_2\inn k)^n+(k_3\inn k)^n(k_2\inn k)^m\bigg]
[k_{3a}u'(2\xi.k_{4})+k_{4a}r'(2\xi.k_{3})
-\xi_a u' r']
\ ,\eeqa
where $k^a$ is the off-shell gauge field momentum. 
The above vertices yield the following infinite set of massless poles in the EFT:
 \beqa
8i\mu_p \eps^{a_{0}\cdots a_{p-1}a} H_{a_{0}\cdots a_{p-3}} k_{1a_{p-2}}
\xi_{1a_{p-1}}
 \frac{1}{(p-2)!(u'+r'+w')}
\sum_{n,m=0}^{\infty}(a_{n,m}-b_{n,m})\label{88}\\ (u'^mr'^n+u'^n r'^m) 
[k_{3a}u'(2\xi.k_{4})+k_{4a}r'(2\xi.k_{3})
-\xi_a u' r']\ ,
\nonumber\eeqa
which indeed agree with the massless singularities of the corresponding string amplitude \reef{masslesspole}; for example, in the leading order, that is, for $ n = m = 0$, the string amplitude produces $4d_{0,0}u'r'$, while at zeroth order in $\alpha'$ the EFT yields
\beqa
16 u' r'(a_{0,0} - b_{0,0}) &=& \frac{-4\pi^2}{3} u' r'\ ,
\eeqa
which is identical to $4d_{0,0}u'r'$.
At first order of $\alpha'$, the string amplitude yields 
$4d_{1,0}u'r'(u'+r')$, and in the modified EFT we obtain
 \beqa
8 u' r' (u'+r')(a_{1,0} + a_{0,1}-b_{1,0} -b_{0,1}) &=& 32\xi(3)u'r'(u'+r')=4d_{1,0}u'r'(u'+r')\ ,
\eeqa
in perfect agreement with worldsheet computations.
One can check that the equivalence holds to all orders.
Hence, all gauge field poles of the D$\overline{\rm D}$ system are reconstructed from the proposal \reef{lagrang} for the all-order two-tachyon-two-gauge field couplings of EFT Lagrangian in an exact and ambiguitiy-free fashion.

In particular, we have found that it is necessary to embed the mixed coupling   
\beqa F^{(1)}\cdot F^{(2)} \eeqa
into the tachyon-extended DBI--SYM action \reef{exp1} to match EFT and string amplitudes.  

\section{Conclusion}

In summary, using worldsheet methods, we have obtained the two-tachyon-two-gauge field couplings in \reef{lagrang} of the EFT on the worldvolume of D$\overline{\rm D}$ systems to all orders in $\alpha'$.
%

Working perturbatively, whereby all kinetic terms remain uncorrected (to leading order in $g_s$), the massless and tachyonic singularities shed light on the structure of higher-derivative corrections to couplings, and help in determining the exact coefficients of the aforementioned couplings. 
In particular, we have found that already at the leading order, the DBI-SYM action \reef{exp1} contains $F^{(1)}F^{(2)}$-couplings in order for the EFT to be consistent with the worldsheet\footnote{In previous works, it has been shown that $D\phi^{i(1)}\cdot D\phi_{i(2)}$-couplings are required on the worldvolume of D$\overline{\rm D}$ systems for consistency between EFT and worldsheet descriptions \cite{Hatefi:2016yhb}.}. 
We have also found that the $C_{p-1}\wedge F$ coupling of the WZ term does not receive any higher-derivative correction.

Let us comment on how our results may provide a guide along future directions towards new ideas in the context of non-BPS branes.

Note that if one uses the ordinary trace  
effective action \cite{Sen:2003tm} instead of \reef{nonab}, the resulting effective action cannot generate all-order singularities of the string amplitudes. 

We note that the tachyon potential that we used is the same as in boundary string field theory (BSFT), \ie
\beqa
V( T^iT^i)&=&1+\pi\alpha'm^2{ T^iT^i}+
\frac{1}{2}(\pi\alpha'm^2{ T^iT^i})^2+\cdots\ ,
\eeqa  
where $m_{T}^2=-1/(2\alpha')$ and $T_{p}$ is the tension of a $p$-brane. 
The expansion complies with the first couple of terms of the potential $V(|T|)=e^{\pi\alpha'm^2|T|^2}$ taken from BSFT \cite{Kutasov:2000aq,Hatefi:2012cp}.
 
If one just uses the derived effective actions of this paper, one can precisely generate all infinite singularities and contact terms of string theory.
Indeed we also discussed that the presence of new mixed couplings like $F^{(1)}\cdot{F^{(2)}}$, as well as $D\phi^{(1)}\cdot{D\phi^{(2)}}$, are inevitable and necessary to derive the actual and consistent results that are demanded when matching string computations with EFT.  

Combining EFT techniques with stringy symmetries \cite{Schwarz:2013wra}, whereby the S-matrix must be symmetric under exchanging $s$ and $t$, and have massless $u$-channel poles, one learns that all $k_i.k_j \rightarrow 0$, and hence we revealed that the unique expansion $(u\rightarrow 0,\, s,t\rightarrow -1/4)$ for the five point function.
 
Carrying out the symmetrized trace in DBI action over $\sigma$ factors, one finds
\beqa
\frac{1}{2}{\rm TrS}\left(V({ T^iT^i})\sqrt{1+[T^i,T^j][T^j,T^i]}\right)&=&\left(1-\frac{\pi}{2}T^2+\frac{\pi^2}{24}T^4+\cdots\right)\left(1+T^4+\cdots\right)\ .\eeqa
Hence, the tachyon  gets condensated at $T\rightarrow \infty$, thereby the tachyon potential tends to zero precisely.  

It is natural to think of the all-order highly symmetric EFTs as counterparts to the holographic duals of Vasiliev's higher spin gravitys \cite{Vasiliev:2012vf} in their turn thought of as consistent truncations of FCS models; for a review, see \cite{Arias:2016ajh}.
The latter are formulated in terms of horizontal Quillen superconnections on  noncommutative fibered manifolds with star-product local FCS action functionals in which all semi-classical nonlocalities have been incorporated into differential graded algebras.
The differential graded algebras serve as natural configuration spaces for noncommutative sigma models of Alexandrov--Kontsevich--Schwarz--Zaboronsky (AKSZ) type \cite{Boulanger:2012bj}; Vasiliev's higher spin geometries subject to boundary conditions then arise on-shell as strict operator realizations of these algebras in representations.

The FCS models contain defects of co-dimension one in their turn containing different sub-defects in various co-dimensions; thus, the FCS models serve as ``grandparents'' for cascades of noncommutative gauge theories of AKSZ type.
It is natural to expect these cascades to induce chains of dualities of topological field theory type between sub-algebras inhabiting the sub-defects, treated as objects of a geometric category, connected by algebraic morphisms mediated by the theory inhabitating the parent defect, treated as morphisms of the geometric category.

In these schemes, Vasiliev's theory plays the role of a parent theory arising in the FCS model in co-dimension one in its turn containing different sub-defects in various co-subdimensions.
In particular, in co-dimension there are sub-defects harbouring fractional spin gravity given by extensions of the original\footnote{The motivation behind \cite{Boulanger:2013naa} was to construct field theories describing the coupling of gravity and gauge fields in three dimensions to fractional spin fields; indeed, the unfolded conformal matter fields sit in infinite-dimensional representations whose spins, viewed as labels of the conformal group $SO(2,3)$, can be deformed continuously within the higher spin extension, while the deformation breaks of $SO(2,3)$ down to $SO(2,2)$.} fractional spin gravity of Chern--Simons type \cite{Boulanger:2013naa} by unfolded conformal matter fields \cite{Vasiliev:2012vf,Nilsson:2013tva,Nilsson:2015pua} arising in the noncommutative AKSZ sigma model in zero- and two-forms in bi-fundamental representations of conformal higher spin and color groups with topological gauge fields\footnote{As shown in \cite{Boulanger:2013naa}, color gauge fields are crucial for Cartan integrability; under extra assumptions the color group can be reduced to $U(1)$, suggesting that an extra gauge field is required in order to complete the Noether procedures set up in \cite{Vasiliev:2012vf,Nilsson:2013tva,Nilsson:2015pua} to all orders.}.

Thinking of the FCS model as a tensionless analog of an open-closed string field theory, with Vasiliev's higher spin gravity and fractional spin gravity, respectively, playing role of closed- and open-string sectors, the counterpart of open-closed string duality in the case of a single stack of D-branes becomes the refined holography proposal involving developed largely independently by Vasiliev \cite{Vasiliev:2012vf} and Nilsson \cite{Nilsson:2013tva,Nilsson:2015pua}, thus representing a morphism from a single sub-defect to the empty set.
By this reasoning, a simple proposal for a tensionless analog of a D$\overline{\rm D}$ system is thus a higher spin gravity model with two fractional spin gravity duals, thus representing a morphism from between two sub-defect.

The derivative structure of cubic and quartic vertices of the D$\overline{\rm D}$ EFT and of effective Fronsdal actions for higher spin gravity are similar.
Indeed, the cubic vertices only have a finite number of derivatives for a given set of spins, while the quartic vertices are given by all-order derivative expansions weighted by $\alpha'$ and the cosmological constant in the D$\overline{\rm D}$ EFT and higher spin gravity, respectively.
One may thus propose that there exists a geometrization of the multi-linear Moyal-like couplings of the EFT at high energies by extending its Quillen superconnection to that of a suitable fractional spin gravity theory (including propagating SYM fields in its matter sector), and correspondingly switching from the metric-like DBI-SYM plus WZ action to a dual noncommutative AKSZ action. 
The resulting configuration spaces of strictly quantized differential graded algebras could provide a framework for resolving classical singularities in SYM gauge theory; for related results in higher spin gravity, see \cite{Aros:2019pgj}.

\section*{Acknowledgments}

EH  would like to thank A. Sagnotti, L. Alvarez-Gaume, R. Antonelli, A. Kuntz, K. Narain, N. Arkani-Hamed, J. Schwarz, A. Sen and  W. Siegel for  fruitful discussions.   This work is supported by INFN (ISCSN4-GSS-PI), by Scuola Normale Superiore, and by MIUR-PRIN contract 2017CC72MK003.
PS would like to thank C. Arias, R. Aros, N. Boulanger, F. Diaz, E. Sezgin, M. Valenzuela and B. Vallilo for discussions.


\end{document}